\documentclass[
   final            
  ]
  {aipproc}

\layoutstyle{6x9}

\def\slash#1{#1 \hskip-0.50em /}
\def\Slash#1{#1 \hskip-0.65em /}

\begin{document}
\title{Short- and long-distance QCD effects\\ in $B$\/-meson decays}

\author{Th.~Feldmann}{
  address={CERN Theory Division, CH-1211 Geneva 23, Switzerland}
}

\begin{abstract}
Various exclusive and inclusive decays of $B$~mesons are studied,
at present, with dedicated experiments at ``$B$~factories''. 
In order to compete with the experimental accuracy, 
we need a reliable theoretical framework to compute
strong interaction effects in a hadronic environment. 
I discuss how the separation of (perturbatively calculable)
short-distance QCD effects from (non-perturbative)
long-distance phenomena 
helps to obtain precise theoretical predictions.

\vspace{1em}

\hfill {{\tt [nucl-th/0310008, CERN-TH/2003-236]}}
\end{abstract}

\maketitle


\section{Introduction}

The standard model (SM) for electroweak and strong interactions
of elementary particles allows for precise 
theoretical predictions that are in 
remarkable agreement with present experimental results
from high-energy experiments. However, an explanation of the
SM and its parameters within a more fundamental theoretical
framework remains an unresolved puzzle. 
Many questions in this respect 
(e.g.\ a grand unification
of forces and matter, the origin of neutrino masses and
its relation to the quark sector, the origin of
$CP$ violation, etc.)
are related to the flavour sector of the SM, i.e.\
to the masses and coupling constants of the different quark
and lepton species. 
$B$\/-meson decays are particularly useful
to explore the parameters of the flavour sector in the SM
or possible extensions of it. The $b$ quark has a relatively long
lifetime, and provides many rare decay modes that have small
branching ratios and are sensitive to details
of the interactions at small distances. In this way one hopes
to reveal indirect effects from physics beyond the SM 
even before the direct detection of new particles.

From the QCD point of view, $b$ quarks are interesting 
because they are the heaviest
quarks that build pronounced hadronic bound states.
The fact that the $b$\/-quark mass  ($m_b \simeq 5$~GeV)
is large with respect to typical hadronic scales leads to new
approximate symmetries that can be observed in the $b$\/-hadron
spectrum and decays.
Furthermore, the heavy-quark mass provides a scale at which
the strong coupling constant is still small, and perturbative
computations of short-distance effects are possible. 
The short-distance dynamics 
has to be separated from (non-perturbative) 
long-distance physics related to the intrinsic
QCD scale $\Lambda_{\rm QCD}$.
Technically this is achieved by an operator product expansion
and from an effective field theory approach, respectively.
In this way, $B$\/-meson decays can also be used to test 
and improve our
theoretical understanding of QCD in different dynamical regimes.


\section{QCD at the electroweak scale}

\begin{figure}[tbptb]
\begin{tabular}{lll}
{\includegraphics[width=0.28\textwidth,bb = 120 640 295 755]{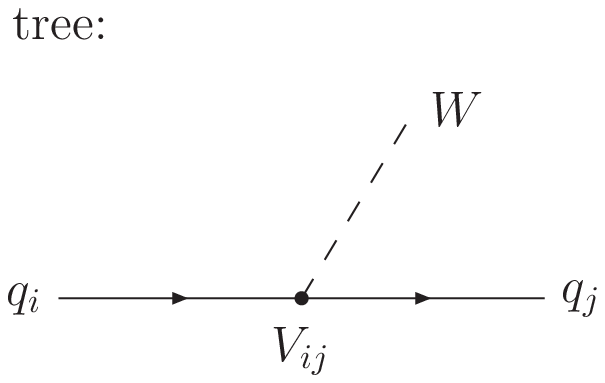}} & 
{\includegraphics[width=0.28\textwidth,bb = 120 640 295 755]{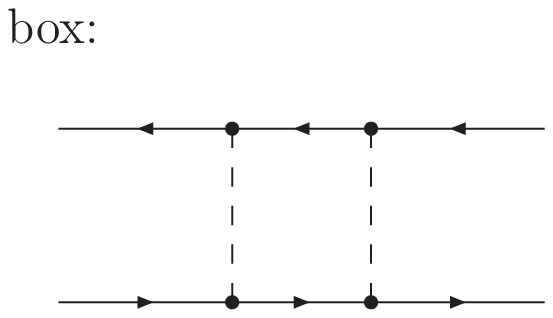}} & 
{\includegraphics[width=0.28\textwidth,bb = 120 640 295 755]{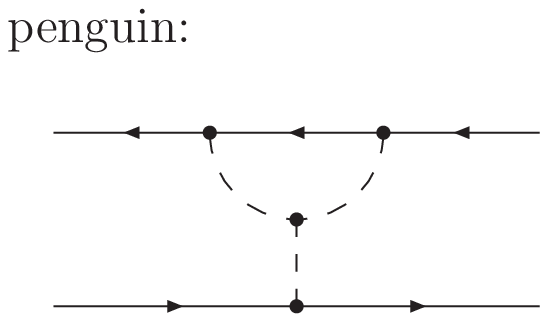}}  
\end{tabular}
 \caption{Weak flavour transitions in the standard model.}
\label{topfig}
\end{figure}

In the SM, transitions between different 
quark flavours are mediated by charged $W$ bosons. The relative
strengths $V_{ij}$ for flavour transitions $q_i \to q_j$ 
define the unitary Cabbibo--Kobayashi--Maskawa (CKM)
matrix,  which can be parametrized in terms of 3 real angles
and one $CP$\/-violating phase.
On the other hand,
flavour-changing {\em neutral}\/ currents (FCNCs) can only
be induced via loop diagrams (box or penguin topologies,
see Fig.~\ref{topfig}). As a consequence,
FCNCs are sensitive to the properties of virtual heavy
particles in the loops, e.g.\ the top quark
in the SM, or new particles in extensions
of it. 

\subsection{Effective Hamiltonian}

As in the case of the muon decay, we may 
``integrate out'' the heavy particles ($W,Z$~bosons, top quark,
new physics) to arrive at an effective Hamiltonian (for a specific
flavour transition), which has the schematic form
\begin{equation}
   H_{\rm eff} \ \propto \ G_F  \, \sum\limits_i C_i(\mu) \cdot {\cal O}_i
\ + \ \mbox{\small terms suppressed by $1/m_W^2$.}
\label{Heff}
\end{equation}
Here $G_F$ is the Fermi constant, and $C_i(\mu)$ are effective
coupling constants (Wilson coefficients)
that encode the short-distance 
dynamics from physics {\em above}\/ the scale $\mu = {\it O}(m_W)$.
The dynamics of the
remaining five quark flavours, leptons, gluons and photons
at energy scales {\em below}\/ $\mu$ is described by a set
of operators ${\cal O}_i$ (four-fermion operators, and
operators that couple fermions
to the chromomagnetic or electromagnetic field strength).
For more details, see the review in \cite{Buchalla:1996vs}.

\subsubsection{Matching}

\begin{figure}[t]
\begin{tabular}{lcl}
\parbox[c]{0.38\textwidth}{\includegraphics[width=0.32\textwidth]{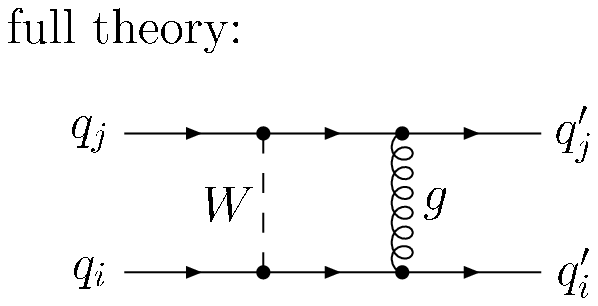} }
 &  $\leftrightarrow$ \ \ &
\parbox[c]{0.35\textwidth}{\includegraphics[width=0.24\textwidth]{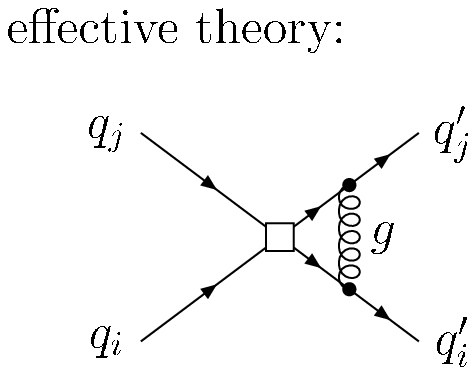}} 
\\[-0.5em] &&
\end{tabular}
\caption{Example for radiative correction in the 
 full and in the effective theory.}
\label{match}
\end{figure}

The Wilson coefficients can be calculated by ``matching'' 
scattering amplitudes in the SM and the effective theory. 
This requires computation of  QCD (and QED) radiative corrections 
in both cases; see Fig.~\ref{match}.
Since the strong coupling constant
is small at the matching scale, the Wilson coefficients
have a perturbative expansion,
\begin{equation}
  C_i(m_W) = c_i^{(0)} + \frac{\alpha_s(m_W)}{4\pi} \, c_i^{(1)} + \ldots
\end{equation}
At present, the SM matching coefficients are known at order
$\alpha_s^2$ (i.e.\ at two loops).
Different models for physics at and above the electroweak scale
yield different matching coefficients:
\begin{equation}
  c_i^{(n)} = c_i^{(n)}(m_t,m_W,\ldots) + c_i^{(n)}(\mbox{new physics}) \ .
\end{equation}
Therefore, experimental measurements of Wilson coefficients 
in weak decays test the SM and/or constrain new physics models.

\subsubsection{Resummation of large logarithms}

Radiative corrections to the matrix elements of the 
effective operators ${\cal O}_i$
in general would involve large logarithms $\ln m_b/\mu$,
when calculated at a scale $\mu \sim m_W$. 
Since higher orders in perturbation theory also 
lead to higher powers of logarithms, the convergence of the
perturbative series would be poor.
This can be avoided by exploiting the fact that 
Wilson coefficients in the effective theory
obey a renormalization-group equation,
which allows the evolution of $C_i(m_W)$ to $C_i(m_b)$.
At the low scale $m_b$, 
the large logarithms $(\ln m_b/m_W)^n$ are explicitly resummed in
$C_i(m_b)$, and matrix elements of ${\cal O}_i$ only contain
dynamics from energy scales below $\mu = m_b$.
In order to derive the evolution equations,
one has to calculate the anomalous-dimension
matrix $\gamma_{ij}$ that describes the scale dependence and mixing
of operators in the effective theory. Currently,
the calculation of three-loop anomalous dimensions is being
completed (for a recent contribution, see \cite{Gambino:2003zm}).

\paragraph{Example: $B \to X_s \gamma$}

A prominent phenomenological
example, where the theoretical machinery of the
effective-Hamiltonian approach is relevant, is the
inclusive rare radiative decay $B \to X_s\gamma$. 
It provides a stringent test of the contributions to the
Wilson coefficient  $C_7$, which is related to the effective $b \to s\gamma$
vertex. The comparison between experiment and theory reads
\begin{eqnarray}
\mbox{Exp.:} && {\rm BR}[B \to X_s\gamma] = (3.34 \pm 0.38)\times 10^{-4} 
\qquad \mbox{\cite{Jessop:2002ha}} \\
\mbox{SM:} && {\rm BR}[B \to X_s\gamma] = (3.70 \pm 0.30)\times 10^{-4}
\qquad \mbox{\cite{Gambino:2001ew,Buras:2002tp}}
\label{Xsg}
\end{eqnarray}
The theoretical uncertainty is dominated by the
renormalization-scheme dependence induced by the charm quark mass.
The experimental and theoretical accuracy is already sufficient
to put strong constraints on many new physics models
(for a recent review on inclusive rare $B$ decays, 
see \cite{Hurth:2003vb}).


\section{QCD and heavy-quark expansion}

Because quarks are confined, experiments can only probe weak
interactions in a hadronic environment.
Technically, this amounts to considering hadronic matrix elements
of the effective Hamiltonian \eqref{Heff}. By construction,
these matrix elements are sensitive to long-distance QCD dynamics,
which is not accessible in perturbation theory. Nevertheless, 
some simplifications arise from the fact that the $b$ quark
mass is large compared to $\Lambda_{\rm QCD}$.
On the one hand, the strong coupling constant is small,
$\alpha_s(m_b) \ll 1$,
which implies that the dynamics at distances of order $1/m_b$ is
still perturbative. On the other hand $\Lambda_{\rm QCD}/m_b \ll 1$ 
provides a
small expansion parameter, and in the heavy-quark limit
($m_b \to \infty$) the
number of independent unknown hadronic quantities may 
be less than in the general case.

\subsection{Heavy quark effective theory (HQET)}

The above observations can be formalized in terms of
an effective theory (HQET) where -- to first approximation --
heavy $b$ and $c$ quarks are replaced by static colour/flavour sources,
moving with a fixed velocity $v^\mu$,
\begin{equation}
  h_v(x) = e^{i m_Q \, v\cdot x} \, \frac{1+\slash v}{2} \, Q(x) \ .
\label{hv}
\end{equation}
The first few terms in the
effective Lagrangian, resulting from integrating out
``small'' spinor components and ``hard'' 
quark and gluon modes (with virtualities of order $m_Q^2$),
reads
\begin{equation}
 {\cal L}_{\rm HQET} 
      = \bar h_v  \left\{ i v \cdot D +
               \frac{(i \vec D)^2}{2m_Q} +
               C_m(\mu) \, \frac{g_s}{4m_Q} \, \sigma_{\mu\nu} \, G^{\mu\nu} + \ldots  \right\} h_v \ ,
\label{Lhqet}
\end{equation}
(where $v \cdot \vec D = 0$).
The first term in this expansion is independent of the heavy-quark mass
and diagonal in the heavy-quark spin. As a consequence, two new
symmetries arise in the heavy-quark limit: heavy-flavour symmetry
reflects the fact that the soft interactions of the $b$ or $c$ quark
become the same, once the center-of-mass motion of the heavy quark
is subtracted. Heavy-quark spin symmetry is related to the fact that
soft interactions do not change the spin of the heavy quark.
This has important implications for phenomenological observables
such as the heavy hadron mass spectrum, or heavy meson
transition form factors, to be discussed below.\footnote{HQET also
applies to heavy-to-light decays as long as the energy transfer to
the light quarks and gluons is small. In this kinematic regime,
exclusive heavy-to-light processes are described by an effective
low-energy Lagrangian for heavy and light mesons that combines 
HQET and chiral perturbation theory 
\cite{Wise:1992hn}.
}
The symmetries are broken by the subleading (kinetic and chromomagnetic)
terms in the Lagrangian \eqref{Lhqet},
as well as by perturbative matching coefficients for decay currents. 
For more details and references to the original literature,
see for instance the review \cite{Neubert:1994mb}.

\paragraph{Example: Extracting $|V_{cb}|$ and $|V_{ub}|$}

A well-known application of the heavy-quark mass expansion and
HQET is the determination of the
CKM elements $|V_{cb}|$ and $|V_{ub}|$ from $b \to c$ and
$b \to u$ decays. At leading order in the $1/m_b$ expansion,
the inclusive rates are just given by the partonic subprocess,
which is calculable in perturbation theory
(this has also been exploited in obtaining the theoretical prediction
for $B \to X_s\gamma$ in \eqref{Xsg}). Power corrections
to the inclusive $B \to X$ rates start only at order $1/m_b^2$,
and the corresponding non-perturbative parameters 
can be fitted to moments of experimentally measured
inclusive decay spectra.
As a result, one extracts \cite{Schubert}
\begin{eqnarray}
|V_{cb}|_{\rm incl.} &=& 0.0421 \pm 0.0013 \ ,\cr
|V_{ub}|_{\rm incl.} &=& 0.00426 \pm 0.00013 \pm 0.00050 \ .
\qquad 
\label{vcb}
\end{eqnarray} 
The theoretical accuracy is limited by the
control on higher-order corrections in perturbation theory and
in the heavy-quark expansion. For the
extraction of $|V_{ub}|$ a subtle issue is how to separate
the $b \to u \ell \nu$ spectrum from the $b \to c \ell \nu$
background. 

\subsection{Soft-collinear effective theory (SCET)}

\begin{figure}[btph]
  \includegraphics[width=.6\textwidth]{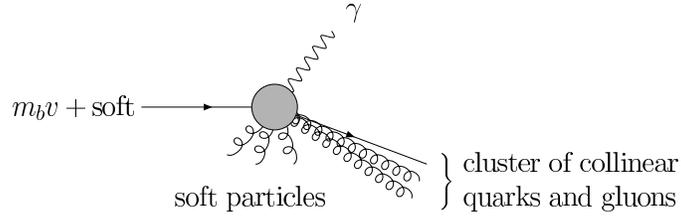}
  \caption{Kinematics for the decay $B \to X \gamma$\/ at large recoil energy.}
\label{setup}
\end{figure}

The heavy-quark expansion can also be
systematically applied to cases
where the heavy quark decays into {\em energetic}\/ light
quarks and/or gluons, see Fig.~\ref{setup}.
Examples are exclusive $B \to \pi\pi$ decays \cite{Beneke:1999br},
or the endpoint spectrum in inclusive $B \to X_s\gamma$ decay
\cite{Bauer:2000ew}.
In addition to the HQET field \eqref{hv} and 
``soft'' (i.e.\ low-energetic) light quark and gluon fields,
the effective theory (SCET) for these cases includes 
``collinear'' quark and gluon modes whose energy is 
proportional to the heavy-quark mass. In particular,
collinear quarks in SCET are described in terms of ``good''
light-cone components $\xi$, 
in terms of which the Lagrangian that describes collinear quarks 
interacting with collinear gluons reads
\begin{equation}
  {\cal L}_{\rm coll} =
  \bar \xi \left\{ i \,  n_- \cdot D
          + (i \Slash D_\perp-m) \, \frac{1}{i \, n_+ \cdot D} \, (i \Slash D_\perp+m) 
\right\} \frac{\slash n_+}{2} \,  \xi \ .
\end{equation}
Here $n_-^\mu$ is a light-like vector, which is
determined by the jet axis or the momentum of the outgoing hadron(s),
and $n_+^\mu$ is another light-like vector with $(n_+  \cdot n_-) = 2$.
Contributions to heavy-to-light decays 
from the ``bad'' spinor components of collinear
quarks, as  well as from 
interactions between soft and collinear particles are 
suppressed by either $1/m_b$ or by $\alpha_s$ 
(see e.g.\ \cite{Bauer:2000yr,Beneke:2002ph}).
In this way SCET provides an elegant alternative to understand
factorization theorems for various QCD processes \cite{Bauer:2002nz}.
Although the physical motivation for separating short- and long-distance
physics in HQET and SCET is similar, there are important differences to
be noted: 
\begin{itemize}
  \item SCET gives no constraints for the light
        hadron spectrum.
  \item Interactions between energetic particles in the final
        state with soft spectators in the
        $B$ meson are mediated by particles with
        virtualities of order $\mu^2=m_b \Lambda_{\rm QCD}$, 
        which corresponds to a second
        short-distance scale in addition to $m_b^2$.
  \item The effective Lagrangian and currents in SCET are 
        non-local.  
\end{itemize}

\paragraph{Example: Forward--backward asymmetry zero in $B \to K^*\ell^+\ell^-$}

The suppression of the
``bad'' spinor components for light quarks in SCET and heavy quarks
in HQET leads to the reduction of
independent form factors for heavy-to-light transitions at 
large recoil \cite{Charles:1998dr}.
As a consequence, to first approximation, 
the position of the forward--backward asymmetry zero 
in the decay spectrum of the exclusive $B \to K^*\ell^+\ell^-$ can be 
predicted in a model-independent way \cite{Burdman:1998mk,Ali:1999mm}
in terms of the SM Wilson coefficients $C_7$ and $C_9$ 
for $b \to s\gamma$ and $b \to s \ell^+\ell^-$ transitions.
First-order radiative corrections have been calculated 
in \cite{Beneke:2001at} and lead to the estimate $q_0^2 = 4.2 \pm 0.6$,
where $q_0^2$ refers to the invariant mass of the lepton pair
at which the forward--backward asymmetry vanishes.


\section{QCD and hadronic effects}

After separating and calculating the
short-distance QCD effects, we are left with 
matrix elements that encode the long-distance dynamics
of quarks and gluons within hadrons. 
Since QCD perturbation theory cannot be applied in this case, 
we  need alternative strategies to determine these hadronic
parameters and obtain theoretical predictions for physical
observables:
\begin{itemize}
\item extract hadronic parameters from one experimental
      observable, and insert them into the prediction for another;
\item estimate them from non-perturbative
      calculations within lattice QCD;
\item estimate them using 
      QCD (light-cone) sum-rule techniques;
\item tune phenomenological models to a subset of 
      hadronic observables, and estimate other observables within   
      that model. 
\end{itemize}
In all these cases, the ultimate challenge is to give 
a reliable prediction of the systematic uncertainties.
For the first option this apparently is a straightforward 
task, but only practicable if there are enough independent
processes where one and the same hadronic quantity enters
(which is the case, for instance, for parton distribution functions).
In the lattice approach one has to deal with several 
(simultaneous) extrapolations: from finite lattice spacing to the continuum
limit ($a \to 0$), from the ``quenched approximation'' 
($n_f=0$) to dynamical quarks,
from quarks fitting on a finite lattice 
to realistic light- and heavy-quark masses.
Here, progress is to be expected from improved lattice actions,
increasing computer power and optimizing numerical algorithms,
as well as from understanding the systematic effects
for dynamical quark simulations better (see also C.~Davies' talk
at this conference).
QCD sum rules are based on parton--hadron duality, which is 
applied to appropriate correlation functions, from which
properties of the lowest contributing resonance 
can be extracted (for an introduction to QCD sum 
rules, see \cite{Colangelo:2000dp}).
The suppression of contributions from
higher resonances is achieved by 
a Borel transformation, and by relating the  
hadronic and the partonic representation of the
spectral function above some threshold $s_0$.
Among others, systematic uncertainties  
arise from varying the Borel parameter
and the threshold parameter within a ``stability window'', and
from power corrections parametrized in terms of quark and gluon
condensates.
Finally, concerning model estimates, the only way to quantify 
systematic uncertainties often is to compare the predictions of
sufficiently many different models. 

In the following I give some examples of 
hadronic parameters relevant to $B$ physics. 
More details on the numerical estimates for various
quantities can be found in \cite{Battaglia:2003in}.

\paragraph{$f_B$ and $B_B$}

Important phenomenological quantities that characterize the
$B$ meson are the decay constant and the $B_q^0$\/--$\bar B_q^0$ 
mixing parameter.
They are defined as
\begin{eqnarray}
\langle 0 | \bar q \, \gamma_\mu \gamma_5 \, b|B(p)\rangle
        = i f_{B_q} \, p_\mu \ ; \qquad
\langle \bar B^0_q | {\cal O}(\Delta B =2) | B_q^0\rangle\rangle
        = \frac{8}{3} \, B_{B_q}(\mu)\,  f_{B_q}^2 \, m_{B_q}^2 \ ,
\end{eqnarray}
respectively, where ${\cal O}(\Delta B=2)$ is a four-quark
operator that induces $\bar b q \leftrightarrow \bar q b$.
Numerical estimates for these quantities are summarized 
in Table~\ref{tab1} (where we considered the renormalization-group 
invariant quantity $\hat B_{B_q}$ for convenience).
The ratio $\xi$ defined in that table plays an important
role in the analysis of the CKM triangle. The asymmetric error in the
lattice estimate arises from the different ways of performing the
chiral extrapolation to realistic light-quark masses, with or
without including ``chiral logs''. We also remark that, in principle,
$B$\/-meson and $D$\/-meson decay constants are related by HQET. However,
at present, both the experimental accuracy in measuring $f_{D}$
and the theoretical understanding of $1/m_c$ corrections are 
insufficient.

\begin{table}[tpb]
\begin{tabular}{l|ccccc}
\hline
  & $f_{B_{u,d}}$ & $f_{B_s}$ &
   $\hat B_{B_d}$ & $B_{B_s}/B_{B_d}$ & 
   $\xi = f_{B_s} \sqrt{\hat B_{B_s}} / f_{B_d} \sqrt{\hat B_{B_d}}$
\\
\hline
Lattice QCD 
& 203(27)~MeV & 238(31)~MeV & 
1.34(12) & 1.00(3) & 1.18(4)(${}^{12}_0$)\\
QCD sum rules & 
208(27)~MeV & 242(29)~MeV &
1.67(23) & $\simeq 1$& \\
\hline
\end{tabular}
\caption{Estimates for $B$\/-meson decay constant and
$B_q^0$\/--$\bar B_q^0$ mixing parameter from lattice QCD and
QCD sum rules (numerical values taken from \cite{Battaglia:2003in}
).}
\label{tab1}
\end{table}

\paragraph{HQET parameters}

In HQET the masses of the ground-state pseudoscalar ($P$)
and vector ($V$) mesons to order $1/m_Q$ accuracy read 
\begin{eqnarray}
\left\{ \begin{array}{c}
    m_P \\
    m_V
\end{array}\right\} 
&=&
 m_Q + \bar \Lambda(m_q) 
+    \frac{1}{2m_Q} \,
\left\{ \begin{array}{l} 
-\lambda_1(m_q)  - 3 \lambda_2(m_q) \\[0.2em]  
-\lambda_1(m_q)  +  (1-\delta_V(m_q)) \, \lambda_2(m_q)
 \end{array}\right\} \ .
\label{PV}
\end{eqnarray}
In the heavy-quark limit, 
pseudoscalar and vector meson belong to the same spin-symmetry
multiplet characterized by a residual mass term $\bar\Lambda(m_q)$
that only depends on the flavour of the light degrees of freedom.
The parameters $\lambda_1$ and $\lambda_2$ are
related to the kinetic
energy and chromomagnetic operator in \eqref{Lhqet}, respectively.\footnote{We included a correction $\delta_V$ arising from
the mixing between different vector mesons obtained from
light degrees of freedom with spin-parity $j^P=1/2^-$
and $j^P=3/2^-$, respectively.
The mixing angle between members of different spin-symmetry
multiplets vanishes in the heavy-quark limit, 
but may be of the order of $10^\circ$ for the two lightest 
axial-vector $D$ mesons 
\cite{Kilian:1992hq}.
}
The parameters
in \eqref{PV} can be determined from combing information from
spectroscopy, lattice, QCD sum rules, and experimental data on inclusive 
$B$\/-meson decays (see \cite{Battaglia:2003in} for a summary of
quantitative results).
Notice that the definition of the quantities on the right-hand side in
\eqref{PV} is renormalization-scheme and -scale dependent.\footnote{
The recent measurement of new
excited $D$ and $D_s$ resonances 
\cite{Aubert:2003fg}
has renewed the interest
in theoretical predictions for heavy-meson spectra \cite{others}. 
The analogous formula to \eqref{PV} for the lowest-lying
heavy scalar ($S$), and axial-vector ($A$) states reads
\begin{eqnarray}
\left\{ \begin{array}{c}
    m_S \\
    m_{A}
\end{array}\right\} 
&=& 
m_Q + \bar \Lambda'(m_q) 
+    \frac{1}{2m_Q} \,
\left\{ \begin{array}{l} 
-\lambda_1'(m_q)  - 3 \lambda_2'(m_q) \\[0.2em]  
-\lambda_1'(m_q)  +  (1-\delta_A(m_q)) \, \lambda_2'(m_q)
 \end{array}\right\} \ .
\label{SA}
\end{eqnarray}
In general, the HQET parameters for the two different
(would-be) spin multiplets in \eqref{PV} and \eqref{SA}
are independent of each
other.
Additional constraints are obtained if one 
Taylor-expands all quantities around the chirally symmetric
limit (i.e.\ vanishing quark condensate,
$\langle \bar qq \rangle \to 0$), and if one {\em assumes}\/
that higher-order terms in that expansion are suppressed
\cite{Bardeen:2003kt,Nowak:2003ra}. 
In this case 
the mass splitting between chiral multiplets
is expected to be
$\bar\Lambda'(m_q) - \bar \Lambda(m_q) \propto
|\langle \bar q q \rangle|$.
The empirical values for this quantity are in fair
agreement with  phenomenological models based on 
spontaneous chiral symmetry breaking 
\cite{Nowak:1993um,Bardeen:1994ae,Ebert:1995tv,Deandrea:1998uz}.
Furthermore,
neglecting contributions proportional to
$\langle \bar q q\rangle$ or $m_q$ in terms that are
already suppressed by $1/m_Q$, 
one expects
\begin{equation}
  \lambda_1(m_q) \simeq \lambda_1'(m_q) \equiv \lambda_1 \ , \quad
  \lambda_2(m_q) \simeq \lambda_2'(m_q) \equiv \lambda_2 \ , \quad
  \delta_V(m_q) \simeq \delta_A(m_q)  \equiv \delta \ . 
\label{Xrel}
\end{equation}
In other words, to first approximation $1/m_Q$ corrections should not
depend on the flavour and parity of the light degrees of freedom,
which appears to be in good agreement with experimental data.
Nevertheless, one has to
keep in mind that neither $1/m_c$ nor $\langle \bar qq\rangle$ are very small
on hadronic scales,
and higher-order terms in (\ref{PV}--\ref{Xrel})
may be non-negligible.
}

\paragraph{Isgur--Wise form factor}

In the heavy-quark limit, as a consequence of heavy-quark flavour
and spin symmetry, all $B \to D^{(*)}$ form factors are given
in terms of one Isgur--Wise function, 
which is normalized at zero recoil 
\cite{Isgur:1989vq}, 
e.g.
\begin{equation}
\langle D|\bar h_{v'} \, \Gamma \, h_v | B \rangle 
        \propto \xi(v \cdot v') \, {\rm tr}\left[ 
        \frac{1+\slash v'}{2} \, \Gamma \, \frac{1+\slash v}{2} \right]
  \ , \qquad \xi(1) = 1 \ .
\label{xi}
\end{equation}
Perturbative QCD corrections to $\xi(1)$ can be calculated in HQET.
Non-perturbative corrections to \eqref{xi} start only at order
$1/m_Q^2$. The estimate thus obtained for the $B \to D^*$ form factor
relevant to the analysis of $|V_{cb}|$ is \cite{Battaglia:2003in}
$$
h_{A1}^{B \to D^*}(1) = \eta_A 
         \left[1
           + \delta_{1/m^2} + \delta_{1/m^3} + \ldots \right]
        = 0.91^{+0.03}_{-0.04} \ .
$$
The knowledge of the form factor enables one to 
extract the matrix element $|V_{cb}|$ from exclusive 
decay modes.
In practice,  one also needs an estimate of the slope $\xi'(1)$
to extrapolate the experimental data to zero recoil.
In this way one obtains
$|V_{cb}|_{\rm excl.} = 0.0402 \pm 0.0020
$  \cite{Schubert},
which is in good agreement with the number obtained 
from inclusive modes \eqref{vcb}.

\paragraph{Heavy-to-light decays}

In order to extract information on heavy-to-light
decays from {\em exclusive}\/ channels, one needs (among others)
the corresponding transition form factors, for instance
to obtain $C_7$ from $B \to K^*\gamma$ or $|V_{ub}|$ from
$B \to \pi(\rho)\ell\nu$. In contrast to the heavy-to-heavy case,
the values of heavy-to-light form factors themselves
are not restricted by a simple symmetry principle. 
Often, one uses a simple parametrization, e.g.\ for
the relevant form factor for $B \to \pi\ell\nu$ decays:
\begin{equation}
  f_+^{B \to \pi}(q^2) = \frac{f_+^{B \to \pi}(0)}{(1-q^2/m_{B^*}^2)\,
  (1- \alpha \, q^2/m_{B^*}^2)} \ ,
\end{equation}
where the first term in the denominator reflects the fact that for
$q^2 \simeq m_b^2$ the form factor is dominated by the nearest 
vector-meson pole, and $f_+^{B \to \pi}(0)=0.2$--$0.3$ and 
$\alpha = 0.3$--$0.6$
are fitted to lattice, QCD sum rules, or model estimates
\cite{Battaglia:2003in}. 
With increasing experimental data on the exclusive decay spectra,
which would also help to test and refine different model estimates,
significant improvement is expected for the 
determination of $|V_{ub}|$ from $B \to \pi\ell\nu$ and 
$B \to \rho\ell\nu$ decays. Currently, the result for $|V_{ub}|$
from exclusive decays is somewhat smaller than the result from
the inclusive analysis (but compatible within the rather large
uncertainties) \cite{Schubert}.

Other exclusive $B$ decays into light mesons require
additional non-perturbative input, as soon as radiative
QCD corrections are considered.
The first systematic treatment of this issue has been discussed
in \cite{Beneke:1999br}
for the case of non-leptonic $B \to \pi\pi$ decays.
It has been shown that (to leading order in the $1/m_b$ expansion and including
first-order $\alpha_s$ corrections) the 
decay amplitude can be factorized into perturbative short-distance 
coefficient functions and non-perturbative matrix elements that
define heavy-to-light transition form factors, or light-cone
distribution amplitudes for $B$ mesons and pions.
Comparison with experimental data from $B \to \pi\pi$ and 
$B \to \pi K$ decays gives additional constraints on the 
CKM triangle (with some uncertainties coming from the parametrization
of the $B$\/-meson distribution amplitude and from estimating
$1/m_b$ corrections).

\section{Summary}

The different energy scales relevant to $B$\/-meson decays
imply that different dynamical aspects of QCD are probed:
at the electroweak scale amplitudes are
sensitive to the parameters
that describe flavour transitions in the Standard Model 
or its possible extensions. Perturbative QCD corrections
are included by matching the full theory onto an effective
Hamiltonian and evolving 
the corresponding Wilson coefficients to scales of the order
of the heavy-quark mass. At this scale, strong interactions
are still perturbative. Furthermore, hadronic matrix elements can be
expanded in inverse powers of the heavy-quark mass.
This is formally described in terms of heavy quark effective theory
(for $b$ decays into charm quarks or low-energetic light quarks and gluons), 
of soft-collinear effective theory 
(for decays into energetic light quarks and gluons).
Finally, the non-perturbative dynamics related to the intrinsic QCD scale is
described in terms of hadronic parameters that have to be extracted from
experiment or estimated from lattice, QCD sum rules or models.
Reducing the theoretical uncertainties from each of the above dynamical
regimes, and comparing the results
with experimental data from $B$ factories, enables
us to test the flavour sector of the Standard Model, to find or
constrain indirect new physics contributions, and to improve our
understanding of perturbative and non-perturbative QCD.


\begin{theacknowledgments}
I would like to thank M.~Nowak and A.~Polosa for interesting discussions
about the $D_s$ resonances,
and T.~Hurth for a careful reading of the manuscript and
helpful comments.
\end{theacknowledgments}

\end{document}